\documentclass[10pt,journal]{IEEEtran}

\usepackage{amssymb}
\usepackage[algoruled,vlined,linesnumbered]{algorithm2e}
\usepackage{comment}
\usepackage{amsthm}

\theoremstyle{definition}

\theoremstyle{remark}

\usepackage{diagbox}
\usepackage{bm}
\usepackage{amsmath}
\usepackage{float}
\usepackage{graphicx}
\usepackage{subfigure}
\usepackage{xcolor}
\usepackage{array}
\usepackage{stfloats}
\usepackage{booktabs}
\usepackage{multirow} 
\usepackage{bbding}
\usepackage{dsfont}

\usepackage{hyperref}

\usepackage[colorinlistoftodos]{todonotes}

\makeatother

\ifCLASSOPTIONcompsoc
  \usepackage[nocompress]{cite}
\else
   \usepackage{cite}
 \fi

\begin{document}
\title{A Learning Aided Flexible Gradient Descent Approach to MISO Beamforming}
\author{Zhixiong~Yang,~\IEEEmembership{}
        Jing-Yuan~Xia,~\IEEEmembership{}
        Junshan~Luo,~\IEEEmembership{}
        Shuanghui~Zhang,~\IEEEmembership{}
        and Deniz~G$\ddot{\text{u}}$nd$\ddot{\text{u}}$z,~\IEEEmembership{Fellow, IEEE}
        
\IEEEcompsocitemizethanks{
\IEEEcompsocthanksitem Zhixiong~Yang, Jing-Yuan~Xia, Junshan~Luo and Shuanghui~Zhang are with College of Electronic Engineering, National University of Defense Technology, Changsha, 410073, China.\protect\
\IEEEcompsocthanksitem Deniz Gündüz is with the Department of Electrical and Electronic Engineering, Imperial College London, SW7 2AZ, UK, and the `Enzo Ferrari' Department of Engineering, University of Modena and Reggio Emilia, Italy.\protect\


E-mail:(j.xia16, d.gunduz)@imperial.ac.uk, yzx21@nudt.edu.cn.


Jing-Yuan Xia and Shuanghui Zhang are the corresponding authors.


This work is supported by National Natural Science Foundation of China, projects 62171448, 61921001, 62131020 and 62022091, by the Natural Science Fund for Young Talents of Hunan Province under grant 2020RC3029, by the European Research Council project BEACON under grant number 677854, and by CHIST-ERA grant CHISTERA-18-SDCDN-001 (funded by EPSRC-EP/T023600/1). 
(Corresponding authors: Jing-Yuan Xia and Shuanghui Zhang.)

 }

}


\IEEEtitleabstractindextext{%


\begin{abstract}

This paper proposes a learning aided gradient descent (LAGD) algorithm to solve the weighted sum rate (WSR) maximization problem for multiple-input single-output (MISO) beamforming.
The proposed LAGD algorithm directly optimizes the transmit precoder through implicit gradient descent based iterations, at each of which the optimization strategy is determined by a neural network, and thus, is dynamic and adaptive.
\textcolor{black}{At each instance of the problem, this network is initialized randomly, and updated throughout the iterative solution process.} Therefore, the LAGD algorithm can be implemented at any signal-to-noise ratio (SNR) and for arbitrary antenna/user numbers, does not require labelled data or training prior to deployment. Numerical results show that the LAGD algorithm can outperform of the well-known WMMSE algorithm as well as other learning-based solutions with a modest computational complexity. Our code is available at \href{https://github.com/XiaGroup/LAGD}{https://github.com/XiaGroup/LAGD}.

\end{abstract}

}

\maketitle

\IEEEdisplaynontitleabstractindextext

\IEEEpeerreviewmaketitle

\section{Introduction}\label{sec:introduction}

\IEEEPARstart{B}{eamforming} plays an essential role in multi-antenna cellular networks. A fundamental and widely studied problem is the downlink beamforming design, where the goal is to maximize the weighted sum rate (WSR) within a total power constraint. The WSR maximization problem is non-convex, and is known to be NP-hard \cite{luo2008dynamic, xu2017waveforming}. Popular solution approaches either adopt convex approximations \cite{peel2005vector, ng2010linear, kibria2013coordinated}, or use alternating minimization (AM) techniques, where each component problem can be solved in closed form \cite{christensen2008weighted, schmidt2009minimum, shi2011iteratively, luo2019joint, zhang2020robust}. Among them, the iterative weighted minimum mean square error (WMMSE) algorithm \cite{shi2011iteratively} is one of the most widely-implemented approaches balancing a good trade-off between performance and computational complexity.

In recent years, deep learning aided data-driven solutions have received significant attention for the solution of the beamforming problem \cite{sun2018learning, xia2019deep, kim2020deep, xia2021meta, zhang2021model}. Generic solutions employ a deep neural network (DNN), known as the black-box model, to solve the WSR maximization problem directly through data-driven optimization. These approaches replace the time-consuming iterative process by well-trained network models, which significantly reduce the computational complexity. For example, Xia et al. \cite{xia2019deep} trained a DNN by supervised learning, which estimates the beamforming matrix by taking the channel gain matrix as the input of the network. These black-box methods can provide significant benefits in terms of the trade-off between the performance and computational complexity depending on the network architecture.
However, they have two main limitations. First, labelled data is generated by the WMMSE algorithm. Therefore, the performance of these end-to-end learning solutions is limited by that of the WMMSE algorithm. Second, the DNN behavior also results in poor interpretability on algorithmic principles and little availability of incorporating expert knowledge, which severely limits practical applications.

More recently, a model-based learning approach, called \emph{deep unfolding} \cite{pellaco2021deep, chowdhury2021unfolding, gao2018enhanced, li2020deep} has achieved significant success in improving both the performance and the model explainability of DNN-based solutions. The common idea of deep unfolding is to map a known iterative algorithm to a DNN, where each iteration of the original iterative algorithm is represented by one layer of the network. In this way, the optimization inspired structure is maintained, and the expert knowledge can be naturally utilized. The work in \cite{pellaco2021deep} proposes a deep-unfolding method to make the step sizes of the iterative WMMSE algorithm trainable to obtain an adaptive trade-off between computational complexity and performance.
The unfolded WMMSE algorithm has made a significant step forward in model explainability, but the performance of unfolding methods in \cite{pellaco2021deep, chowdhury2021unfolding} are still bounded by the WMMSE algorithm. In addition, although the well-trained unfolded network can achieve better efficiency, the generalization ability is relatively poor. Retraining a new network is necessary when the application settings, such as the channel signal-to-noise ratio (SNR), number of users or antennas, vary in real application scenarios.

In this paper, we propose a learning aided gradient descent (LAGD) algorithm to solve the transmit beamforming problem in a multi-user mutiple-input single-output (MISO) communication system. The proposed LAGD algorithm directly optimizes the transmit precoder, instead of converting the WSR problem into an AM-based framework, i.e., as in the WMMSE algorithm. Thus, the matrix inversions, bisection search on the Lagrange multiplier, or iterations over three sets of variables are avoided.
On the other hand, different from vanilla gradient-based solutions that follow fixed, explicit, and handcrafted algorithmic rules, the LAGD algorithm adopts an implicit and learned optimization rule through a neural network, whose parameters are updated by back-propagating the WSR value. This neural network is updated over the iterations with the goal of finding a better transmit precoder to maximize the WSR value. Therefore, the LAGD algorithm achieves an adaptive and dynamic strategy for solving the WSR maximization problem at each iteration. 


The benefits of the proposed LAGD algorithm include the practical feasibility, flexibility, and interpretability. First, the proposed LAGD algorithm is less data-dependent in practical applications than the existing deep-learning based methods \cite{pellaco2021deep, xia2019deep, huang2018unsupervised}, since the training process is unsupervised, and is essentially embedded with the solution process. 
\textcolor{black}{This allows the LAGD algorithm to be implemented directly on solving beamforming problems without a dedicated training stage.} Therefore, the LAGD algorithm can be used as a feasible \textit{plug-and-play} tool for different scenarios, such as different number of users or antennas, or different signal-to-noise-ratio (SNR) values. 
Besides, instead of replacing the whole gradient descent (GD) iterations by a neural network, the LAGD algorithm retains the iterative structure, and merely inserts a neural network-based update rule into the original GD framework.
Hence, the algorithmic principles of the GD algorithm are retained, while the variable update rule is learned on-the-go throughout the iterations of the algorithm. We argue that the proposed approach not only has better explainability, but also provides more flexibility to incorporate physical understanding of a specific problem, such as expert knowledge or various priors. Moreover, the proposed LAGD algorithm is shown to achieve satisfactory performance even with an arbitrary lightweight network, for example, a fully-connected neural network (FNN) with only one hidden layer of 10 units. Therefore, the overall computational complexity of the LAGD algorithm is quite low, and it is easy to implement in practice.
We demonstrate through simulations that, unlike other learning-based solutions, the LAGD algorithm can outperform WMMSE even with a simple shallow network.

\section{Problem Formulation}\label{Problem Formulation}
We consider a multi-user MISO downlink channel. The transmitter has $M$ antennas and serves $N$ single-antenna users.
The signal received at the $i^{th}$ user is given by
\begin{equation}
y_i = \bm{h}_{i}^{H} \bm{v}_i x_i+\sum^{N}_{j=1,j\neq i}\bm{h}_{i}^{H} \bm{v}_j x_j + n_i   ,\label{eq-0}
\end{equation}
where $x_i \sim \mathcal{CN}(0,1)$ denotes the independent data symbols for the $i^{th}$ user, $\bm{v}_i \in \mathbb{C}^{M}$ is the transmit precoder vector of the $i^{th}$ user, $\bm{h_i} \sim \mathcal{CN}(0,\bm{I_M})$ is the channel vector of the $i^{th}$ user, and $n_i \sim \mathcal{CN}(0,\sigma^2)$ denotes the independent additive white Gaussian noise with power $\sigma^2$. We assume that the channel gain vectors $\bm{h}_1,...,\bm{h}_N$ are known at the transmitter and the receivers. The signal-to-interference-plus-noise-ratio (SINR) at the $i^{th}$ user is given by
\begin{equation}
\text{SINR}_i = \frac{\lvert \bm{h}_{i}^{H} \bm{v}_i \rvert^2}{\sum^{N}_{j=1,j\neq i} \lvert \bm{h}_{i}^{H} \bm{v}_j \rvert^2 + \sigma^2} .\label{eq-1}
\end{equation}
The beamforming problem is formulated as the maximization of the WSR subject to a total transmit power constraint, as follows
\begin{equation}
\begin{aligned}
\underset{\bm{V}}{\max} \;F(\bm{V}&) \triangleq \sum ^{N}_{i=1} \alpha_i \log_2(1+\text{SINR}_i) \label{eq-2} \\
    &\text{s.t.}\; \text{Tr}(\bm{V}\bm{V}^H)\leq P, 
\end{aligned}
\end{equation}
where $\alpha_i$ is the weight of the $i^{th}$ user (assumed to be given), $P$ is the maximum total transmit power, $\bm{V}\triangleq[\bm{v}_1,\bm{v}_2 \dots \bm{v}_N]^T$, is the matrix of beamforming vectors, and $\mathtt{Tr}(\cdot)$ denotes the trace operator.

Problem (\ref{eq-2}) is known to be non-convex and NP-hard \cite{luo2008dynamic},
but we can employ a generic GD based solution, in which the variable $\bm{V}_k$ is optimized in an iterative fashion. The update rule at the $k^{th}$ iteration can be written as
\begin{equation}
\bm{V}_{k+1}=\bm{V}_{k}-\gamma_{k}\cdot \text{g}(\nabla_{\bm{V}_{k}}F(\bm{V}_{k})) ,\label{eq-VGD}
\end{equation}
where $\nabla_{\bm{V}_{k}}F(\bm{V}_{k})$ is the gradient of the WSR with respect to current beamforming matrix,
$\text{g}(\cdot)$ denotes a hand-crafted variable update function, and $\gamma_{k}$ represents the step size at the $k^{th}$ iteration. 
However, the GD based solution that follows the iterations in (\ref{eq-VGD}) can be stuck at saddle points or bad local optima, where the gradient vanishes, i.e., $\nabla_{\bm{V}_{k}} F(\bm{V}_{k})=0$. Therefore, the WSR problem is typically solved either by using convex approximation, or in an AM-based framework, such as the WMMSE algorithm \cite{shi2011iteratively}. 
The WMMSE algorithm first converts the original WSR maximization problem (\ref{eq-2}) into an equivalent weighted sum mean square error minimization problem:
\begin{equation}\label{minWSR}
\begin{aligned}
   \min_{\bm{u},\bm{w},\bm{V}}\sum_{i=1}^{N}\alpha_i(w_{i}e_{i}-\log_{2}w_i)\\
   \text{s.t.}\; \text{Tr}(\bm{V}\bm{V}^H)\leq P,
\end{aligned}
\end{equation}
where $e_{i}$ is the mean-square error given by $e_{i}\triangleq|u_{i}\bm{h}_{i}^H\bm{v}_i-1|^2+\left(\sum_{j\not=i,j=1}^N|u_{i}\bm{h}_{i}^H\bm{v}_j|^2\right)+\sigma^2|u_i|^2$,
$u_i$ denotes the receiver gain, $w_i$ is the user weight, $\bm{u}=[u_1,\ldots,u_N]^T$, $\bm{w}=[w_1,\ldots,w_N]^T$.
This problem is convex in each individual variable, and
the WMMSE algorithm iteratively minimizes the objective function with respect to each individual variable by solving these convex optimization problems. However, due to the intrinsic non-convexity of the problem, the AM-based solution can get trapped at bad local optima even though each partial optimization of the individual variables can be solved in closed form.

\section{Proposed LAGD Method}\label{Proposed LAGD Method}
We propose the LAGD algorithm that can learn an adaptive and dynamic iterative updating strategy for optimizing the transmit precoder matrix. At each iteration of the LAGD algorithm, the procoder matrix $\bm{V}$ is updated by a parameterized update function, whose parameters are also updated at each iteration. Specifically, the manually designed variable update function $\text{g}(\cdot)$ in (\ref{eq-VGD}) is replaced by a neural network-based update rule $\text{G}_{\bm{\theta}_{k}}(\cdot)$, where $\bm{\theta}_{k}$ represents the parameters of the network $\text{G}_{\bm{\theta}_{k}}(\cdot)$ at the $k^{th}$ step. 
At the $k^{th}$ iteration, the input to $\text{G}_{\bm{\theta}_{k}}(\cdot)$ is the gradient $\nabla_{\bm{V}_{k}}(F(\bm{V}_{k}))$, which then outputs the term to update the transmit percoder $\bm{V}_{k}$. The formulation can be expressed as
\begin{equation}\label{eq-LAGD-Net} 
    \bm{V}_{k+1}=\bm{V}_{k}+\text{G}_{\bm{\theta}_{k}}(\nabla_{\bm{V}_{k}}F(\bm{V}_{k})).
\end{equation}
In LAGD algorithm, the variable optimization strategy is determined by the network parameters $\bm{\theta}_{k}$,
which are also updated at each iteration by back-propagating the WSR value. 
\textcolor{black}{We would like to emphasize that the network parameters $\bm{\theta}_{k}$ are not trained in advance using a dataset, but instead, they are updated during the iterations of the optimization problem (\ref{minWSR}). 
Accordingly, each update of $\bm{\theta}_{k}$ parameters can be interpreted as \textit{training while solving}, where each previous instance of the problem corresponds to a single training sample.
Essentially, the LAGD algorithm can be used in a plug-and-play fashion, and no training dataset is required.}
The updated network parameters tend to optimize the transmit precoder dynamically over iterations with the goal of maximizing the WSR value.
The parameters of the neural network, $\bm{\theta}_{k}$, are updated by the Adam \cite{kingma2014adam} optimizer:
\begin{equation} \label{eq-6}
    \bm{\theta}_{k+1}=\bm{\theta}_{k}+\alpha_{}\cdot\mathrm{Adam}(\nabla_{\bm{\theta}_{k}} F(\bm{V}_{k+1})),
\end{equation}
where $\alpha_{}$ denotes the learning rate. 

To satisfy the total power constraint $\text{Tr}(\bm{V}\bm{V}^{H})\leq P$, the transmit precoder matrix $\bm{V}$ is projected at each step by
\begin{equation} \label{eq-7}
    \Omega(\bm{V})=\left\{\begin{array}{l}\;\;\;\;\;\bm{V},\;\;\;\;\;\;\;\;\;\;\;\text{if}\;\;\text{Tr}(\bm{V}\bm{V}^{H})\leq P, \\
\frac{\bm{V}}{||\bm{V}||_F}\sqrt{P}, \;\;\;\;\;\;\;\;\;\;\;\text{otherwise}. 
\end{array}\right.
\end{equation} 
The general structure of the proposed LAGD algorithm is presented in \textbf{Algorithm} \ref{alg1}. 


In summary, the LAGD algorithm establishes a trainable variable update function $\text{G}_{\bm{\theta}_{k}}(\cdot)$ to replace the manually designed vanilla function $\text{g}(\cdot)$ in (\ref{eq-VGD}). 
Mathematically, at the $k^{th}$ iteration, the neural network-based learned update rule $\text{G}_{\bm{\theta}_{k}}(\cdot)$ takes the gradient $\nabla_{\bm{V}_{k}}F(\bm{V}_{k})$ as input and predicts the next update term of the transmit precoder. 
The parameters $\bm{\theta}_{k}$ of the update rule $\text{G}_{\bm{\theta}_{k}}(\cdot)$ are also updated over the iterations in order to find a better update function for solving the WSR maximization problem. 
In this way, the algorithmic principles of the original GD-based iterative solution are retained, while the update strategy is endowed with further adaptability and learnability. 

The LAGD algorithm is a novel methodology for solving the non-convex WSR maximization problem. 
In contrast to the vanilla GD algorithm, LAGD seeks to optimize the transmit precoder in a less greedy and more dynamic manner at each iteration, while still identifying the update direction and step size following the same GD principles. This provides the LAGD the capacity to circumvent bad local optima and saddle points on the geometry of the objective function surface.
\textcolor{black}{Note that, thanks to the neural network-based $\text{G}_{\bm{\theta}_{k}}(\cdot)$ in (\ref{eq-LAGD-Net}), non-zero update terms are possible even when the gradient vanishes, i.e., $\nabla_{\bm{V}_{}}F(\bm{V})=0$ \cite{xia2022Metalearning}.}
Compared to unfolding based solutions, the LAGD algorithm also makes a step forward in terms of the interpretability and generalization capabilities. Instead of mapping the iterative algorithm into an end-to-end network model through deep unfolding, LAGD tries to learn only the function-level behavior. Consequently, the iterative variable update process is explainable, and is easy to incorporate expert knowledge and prior information based on physical principles of the problem. 


Next, we highlight the main advantages of the proposed LAGD algorithm:
\begin{itemize}
    \item  Superior performance compared to existing alternatives, including both conventional optimization approaches and the more recent learning-based solutions.
    \item Reduced computational complexity compared to alternative methods thanks to the sufficiency of a lightweight network architecture. The memory cost and computational complexity is consequently highly reduced compared to other DNN-based methods.
    \item Thanks to the unsupervised learning while solving approach and the lightweight network structure, the proposed LAGD algorithm can be used in a plug-and-play fashion in practical applications.
    To be specific, the LAGD algorithm can be implemented to solve beamforming design problems at different SNRs, number of users or antennas at the transmitter, without requiring any prior training procedure.
\end{itemize}

\begin{algorithm}[t]
    \SetAlgoLined
    \textbf{Given:} $F(\bm{V})$, number of users $N$, number of antennas $M$, and channel gains $\bm{h}_1,...,\bm{h}_N$.
    
    \textbf{Initialize:} $\bm{V}_{0}$, $\bm{\theta}_{0}$.
    
    \For{k$\gets$ 0, 1, $\ldots$, K}{
    $\Delta \bm{V} = \text{G}_{\bm{\theta}_{k}}(\nabla_{\bm{V}_k}F(\bm{V}_k))$

    $\bm{V}_{k+1} = \bm{V}_{k} + \Delta \bm{V}$
    
    $\bm{V}_{k+1} = \Omega(\bm{V}_{k+1})$

    $\Delta\bm{\theta} = \alpha \cdot\mathrm{Adam}(\nabla_{\bm{\theta}_{k}}F(\bm{V}_{k+1})) $
    
    $\bm{\theta}_{k+1} = \bm{\theta}_{k} + \Delta\bm{\theta}$
    
    }
    \textbf{Output:} $\bm{V}_{K}$, $F(\bm{V}_{K})$
\caption{\label{alg1}The whole structure of the proposed LAGD algorithm for the WSR maximization problem
}
\end{algorithm}

\section{Simulation results}\label{SIMULATION RESULTS}

In this section, the performance of the proposed LAGD algorithm is evaluated and compared with other alternatives through simulations. The LAGD algorithm is implemented in Python 3.6.13 with Pytorch 1.7.0. The WMMSE algorithm is also implemented in Python 3.6.8 with Tensorflow 1.13.1 for comparison. We assume all the users share the same priority, i.e., $\alpha_i=1$, $\forall i$, while the generalization to non-uniform weights is trivial. The learning rate of the Adam \cite{kingma2014adam} optimizer for the network-based update rule is set to $10^{-4}$. 
\textcolor{black}{While the reported results are obtained by averaging of 1000 realizations of the channel matrix $\bm{H}$ generated independently and identically distributed (i.i.d.) from a complex standard Gaussian distribution, i.e., Rayleigh fading, the proposed LAGD algorithm can be used in any channel distribution.} 


The WMMSE algorithm is applied as the baseline, following the steps in \cite{pellaco2021deep}. We set the maximum number of iterations of WMMSE to 50 (this is set to 6 in previous works) and that of LAGD is set to 500. We randomly initialize the algorithm for 10 times and pick the best results for both LAGD and WMMSE as to limit the negative impact of extremely poor initializations.



\begin{figure}[htbp]
  \centering
  \includegraphics[width=0.8\linewidth]{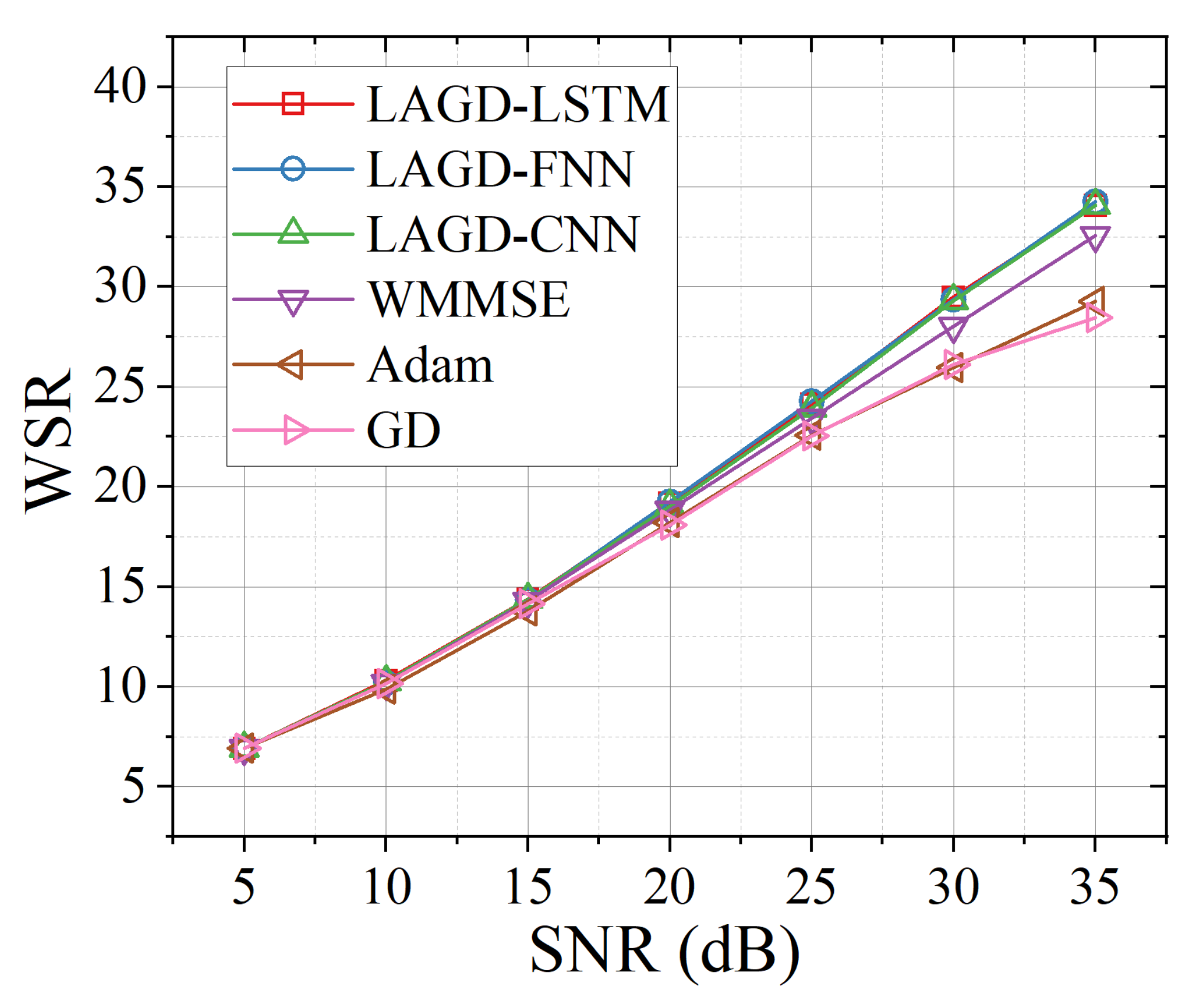}\\
  \caption{\textcolor{black}{LAGD with three network types. Compared with WMMSE and gradient-based conventional GD and Adam schemes. ($N=M=4$).}}\label{SNR-3-type}
\end{figure}

\begin{figure}[htbp]
  \centering
  \includegraphics[width=0.8\linewidth]{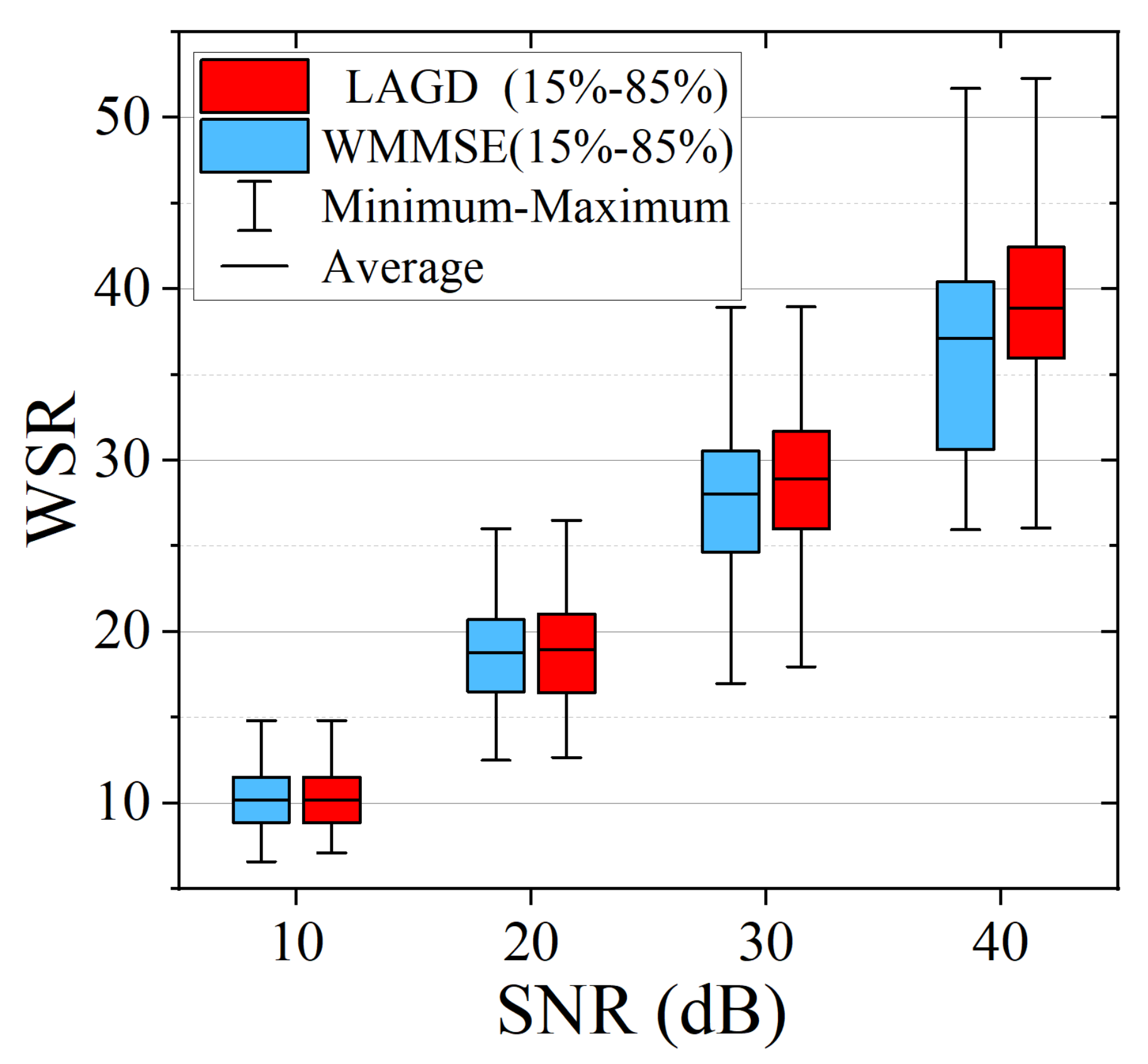}\\
  \caption{\textcolor{black}{The variance of LAGD compared with WMMSE algorithm. ($N=M=4$).}}\label{SNR-variance}
\end{figure}

\textcolor{black}{In Fig. \ref{SNR-3-type}, we evaluate the performance of the LAGD algorithm for different neural network architectures, and compare with the standard GD and Adam \cite{kingma2014adam} (one of the most widely-implemented GD-based algorithms) approaches and the WMMSE performance. Three generic network architectures, FNN, long-short-term-memory (LSTM) and convolutional neural network (CNN) are evaluated for $G_{\bm{\theta}_{k}}(\cdot)$. Specifically, the LSTM network and the FNN contain 2 hidden layers with 40 units in each, while the CNN has 2 convolution layers with one kernel of $3\times3$ in each layer.
The results in Fig. \ref{SNR-3-type} show that all these three networks achieve similar performances, and significantly surpass the WMMSE algorithm in the high SNR regime of SNR$=$20-35dB. 
It can also be noted that the ordinary GD based solutions achieve comparable performance to WMMSE and LAGD when SNR$=$5-20dB. We argue that the non-convexity of the WSR problem grows with SNR. 
Since the behavior is closer to a convex function in the low SNR regime, both WMMSE and conventional GD-based solutions perform reasonably well. However, as the SNR increases, the WMMSE algorithm and ordinary GD-based solutions can be stuck at saddle points or local optima more easily. 
From the superior performance of LAGD in the high SNR regime, we can conclude that its flexible and adaptive update rule allows it to avoid local optima or saddle points.
}

\textcolor{black}{While the results in Fig. \ref{SNR-3-type} are averaged over the channel distribution, in Fig. \ref{SNR-variance}, we further present the variance of the obtained WSR results for the LAGD and WMMSE algorithms over an SNR range from 10dB to 40dB. 
The red rectangles depict the variances of the LAGD results while the blue ones refer to the results from the WMMSE algorithm. 
The number of users and antennas are set to 4, and an LSTM network is used to model function $G(\cdot)$. It is apparent from this figure that the LAGD algorithm also has a smaller variance particularly in the high SNR regime. This is yet another evidence that LAGD can avoid local optima that the WMMSE may get stuck in certain channel realizations.
}  

\begin{figure*}[t]
	\centering
	\hspace*{\fill}
	\subfigure{
		\includegraphics[width=0.23\linewidth]{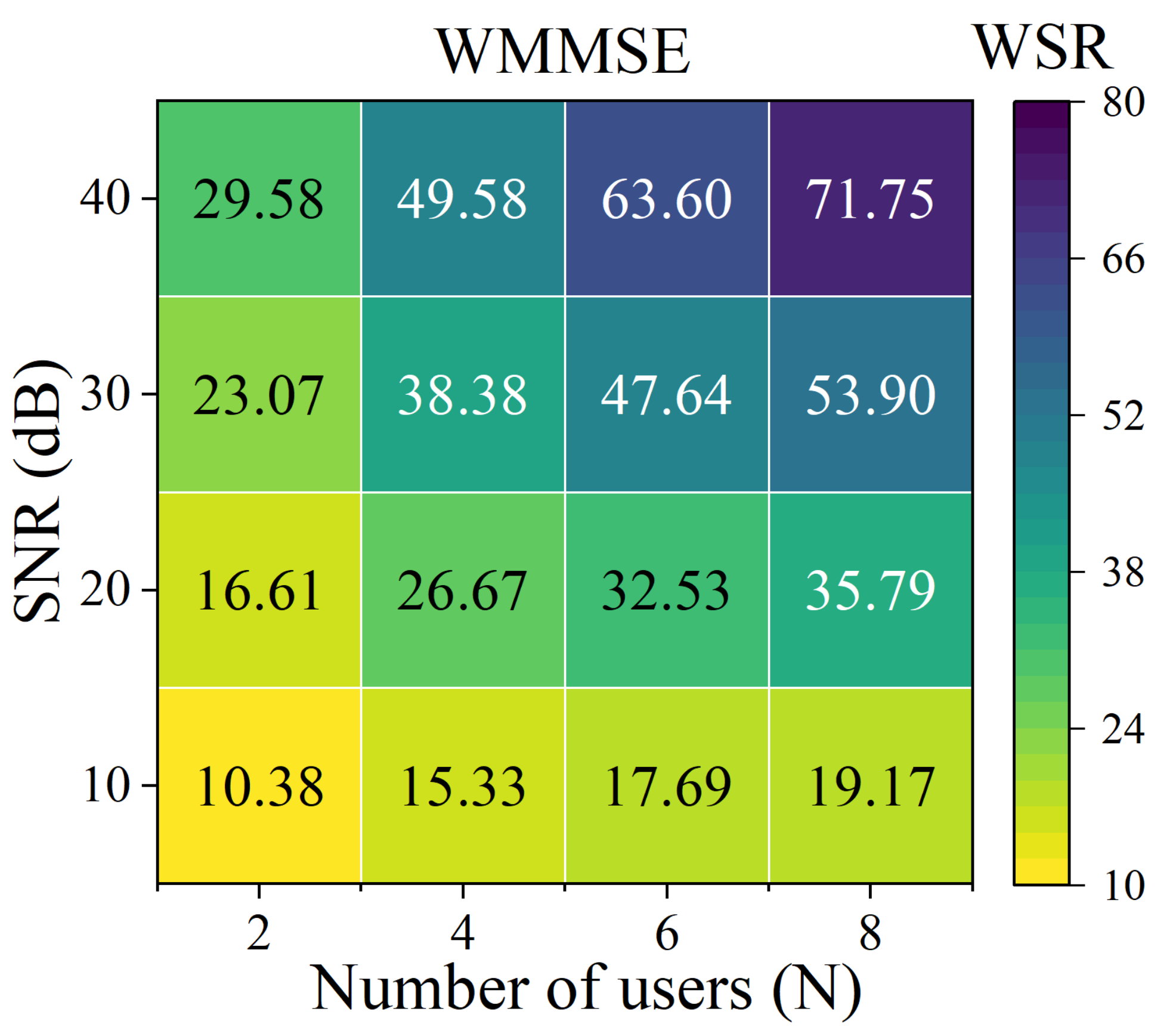}}
		\hfill
	\subfigure{
    	\includegraphics[width=0.23\linewidth]{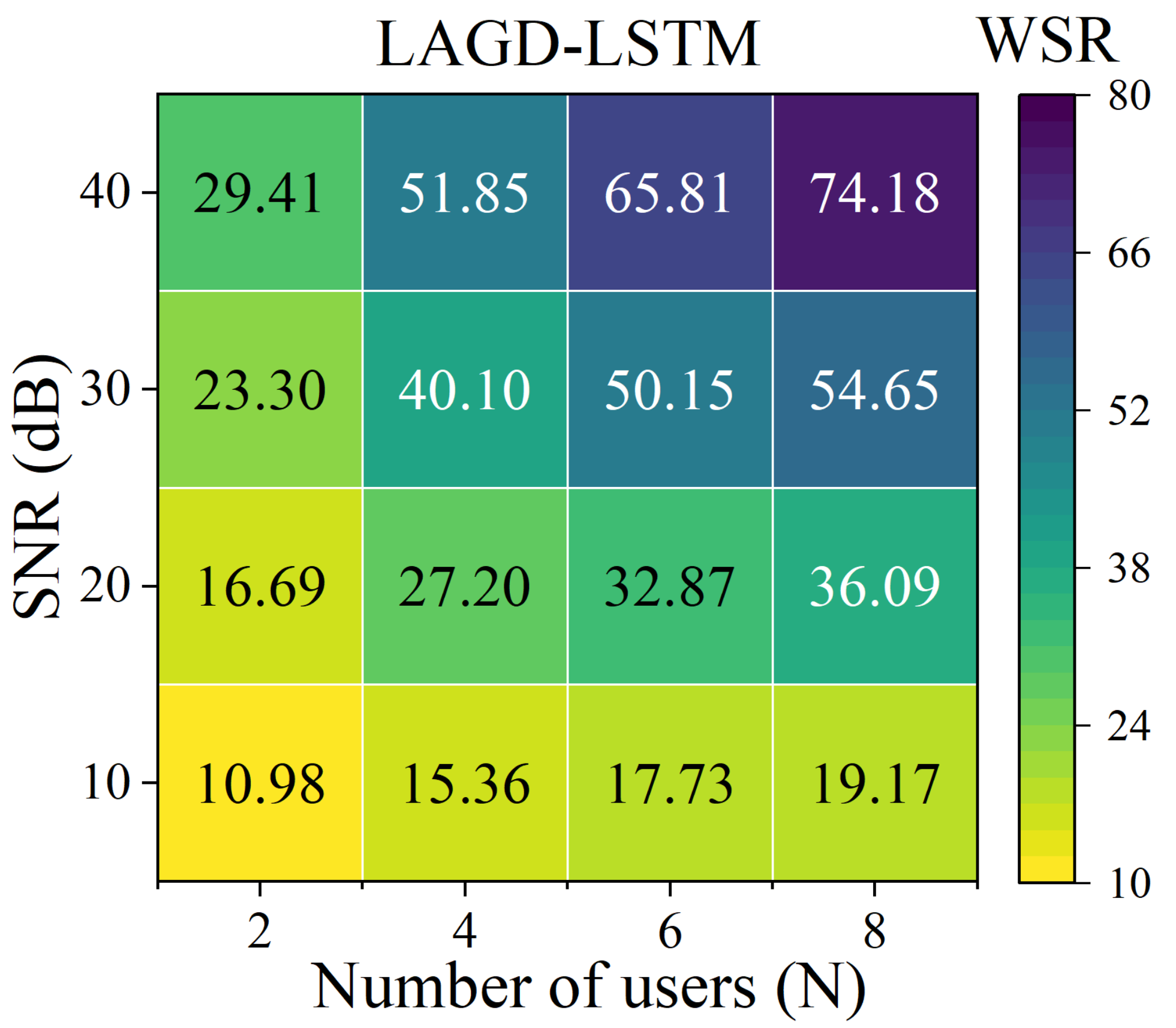}}
    	\hfill
	\subfigure{
    	\includegraphics[width=0.23\linewidth]{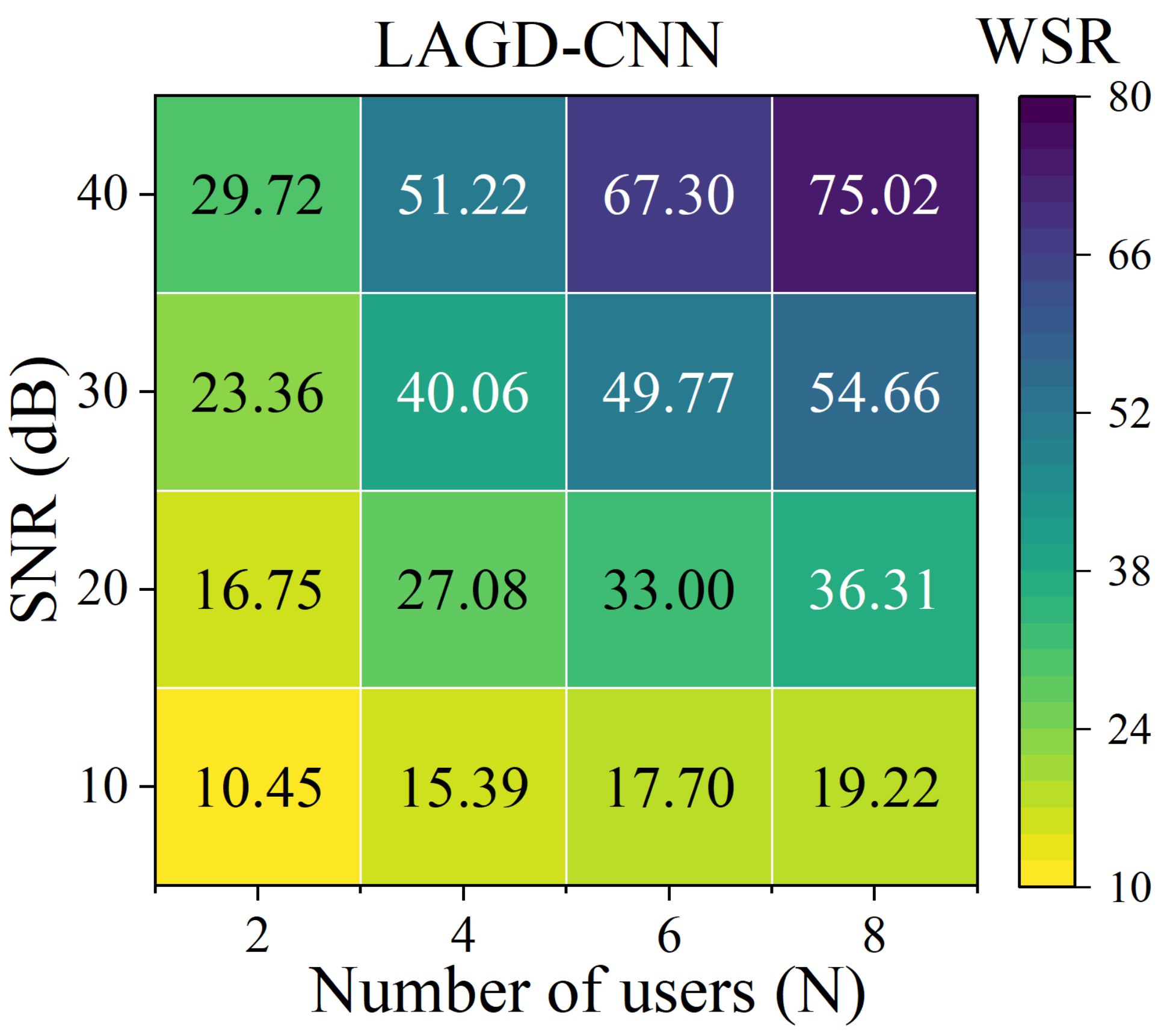}}
    	\hfill
	\subfigure{
		\includegraphics[width=0.23\linewidth]{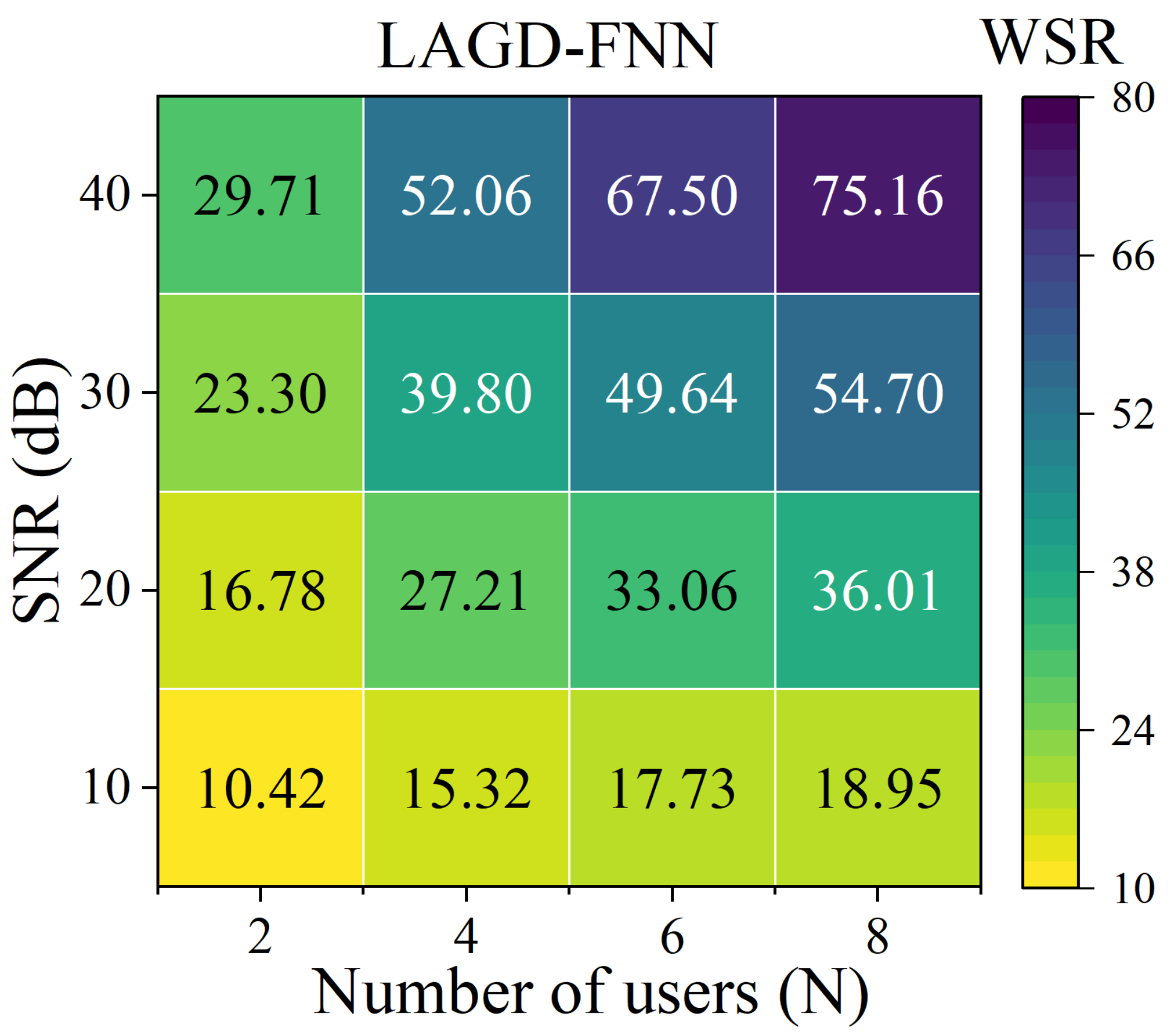}}
	\hspace*{\fill}
	\caption{LAGD and WMMSE performance for different number of users and SNR values.}\label{SNR-USERS-4}
\end{figure*}


\begin{figure}[htbp]
  \centering
  \includegraphics[width=0.8\linewidth]{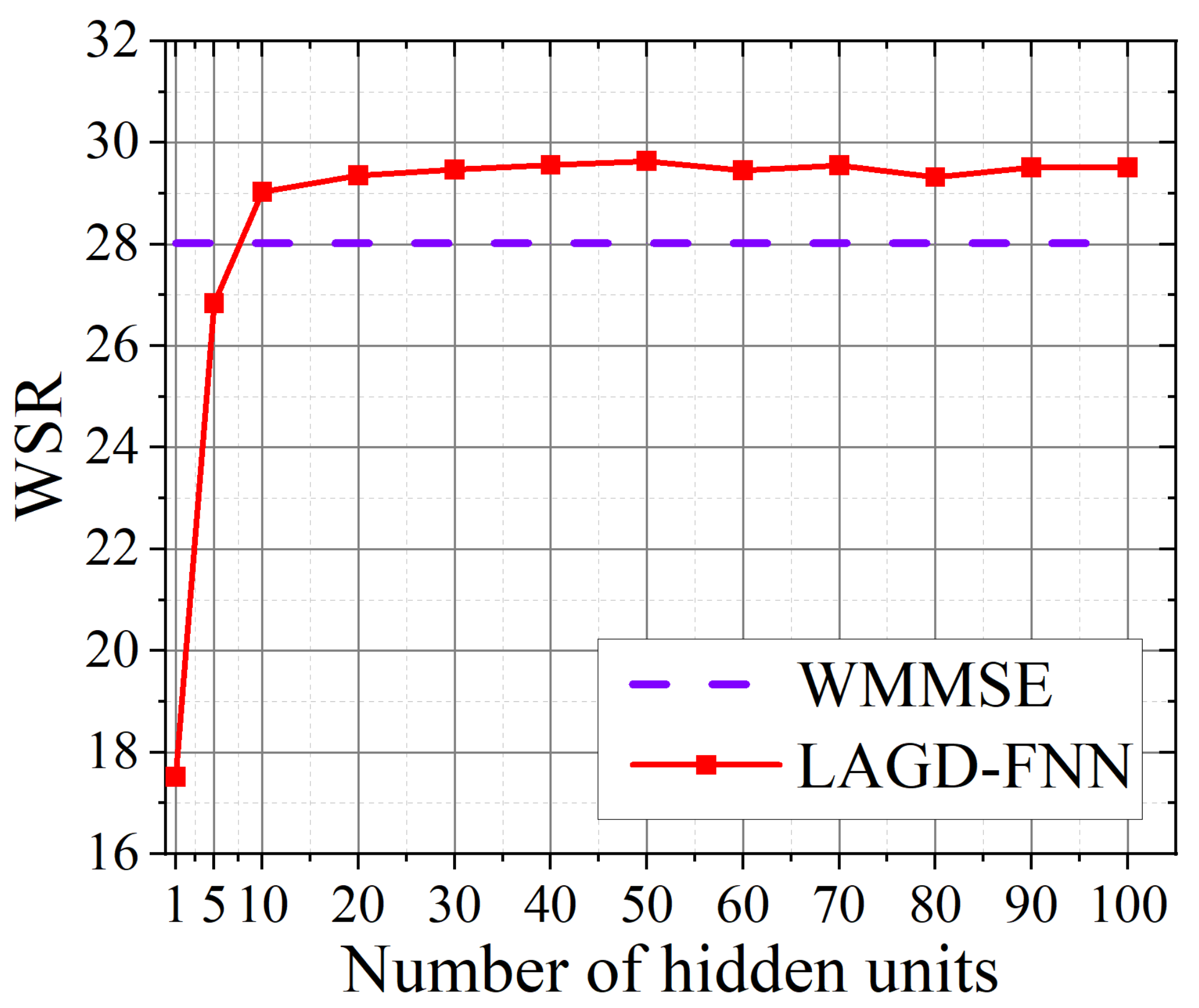}\\
  \caption{\textcolor{black}{WSR obtained with a single hidden layer consisting of different number of units. ($M=N=4$, SNR$=30$dB)}}\label{SNR-UNITS}
\end{figure}

\begin{figure}[htbp]
  \centering
  \includegraphics[width=0.8\linewidth]{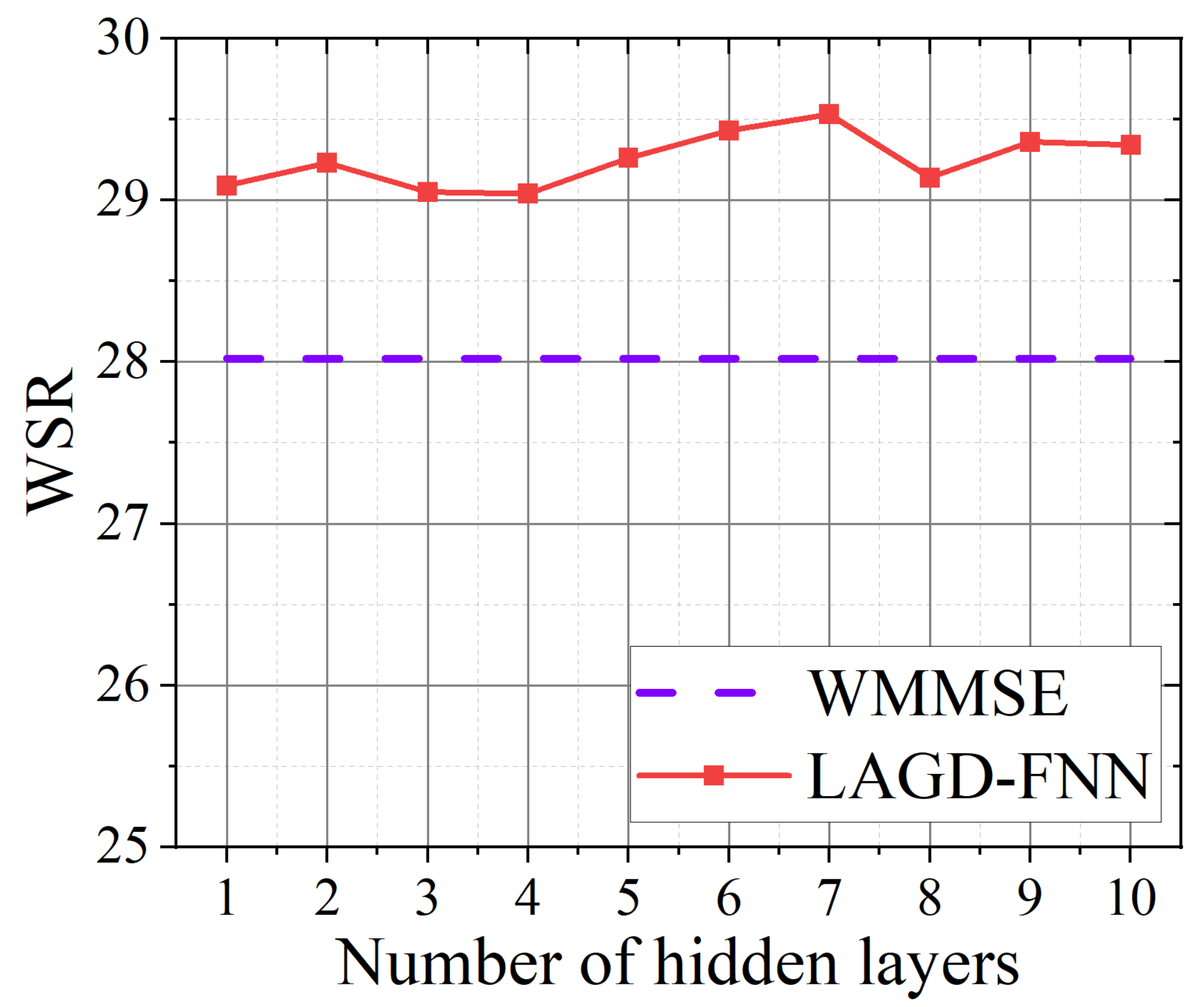}\\
  \caption{\textcolor{black}{WSR obtained with different number of hidden layers and 20 units in each layer. ($M=N=4$, SNR$=30$dB)}}\label{SNR-LAYERS}
\end{figure}

Further simulation results are presented in Fig. \ref{SNR-USERS-4} to demonstrate the generalization capability of the LAGD algorithm when used in different system setups including different combinations of the number of users/antennas and SNR values. The number of users $N$ varies from 2 to 8, while the transmitter has $M=8$ antennas. As before, we consider an SNR range of 10-40dB. We would like to emphasize that, in contrast to the existing deep learning approaches, the LAGD algorithm can be directly implemented in any scenario in a plug-and-play fashion without any prior training procedure when the setting is changed. It can be seen that the WSR values obtained from the LAGD algorithm with three different network architectures show similar behavior to the results presented in Fig. \ref{SNR-3-type}; while the performance improvement is marginal in the low SNR regime, LAGD surpasses WMMSE significantly in the high SNR regime in all the cases. In general, we observe that the improvement with respect to WMMSE increases with the number of users.
This also shows that as the non-convexity of the WSR maximization problem grows with the number of users, the LAGD algorithm becomes even more effective compared to the fixed AM-based strategy of the WMMSE algorithm. The single most striking observation to emerge from the results in Fig. \ref{SNR-USERS-4} is that the FNN architecture reaches a superior performance as the non-convexity of the problem increases, that is when the SNR and the number of users are high. This motivated us to further test the impact of the architecture on the performance. 

In Fig. \ref{SNR-UNITS} and \ref{SNR-LAYERS}, the performance of the LAGD-FNN with different numbers of layers and units are compared. 
We observe that the performance quickly saturates with respect to the number of hidden units. While it can improve by carefully choosing the number of layers, the variations are marginal.
We conclude that the LAGD algorithm does not require a finely-designed DNN architecture to achieve its excellent performance. Using a single-layer FNN to model $G_{\bm{\theta}_{k}}(\cdot)$ dispenses with the cost of network design, and the number of parameters that need to be updated at each iteration is much smaller compared to the existing DNN-based approaches.

\begin{table}[t]  \caption{Comparison of computational requirements} 
 \label{computational complexity}
\renewcommand{\arraystretch}{0.8}
\begin{center}
 \begin{tabular}{llll}  
\toprule   
  Methods                        &Training     &Test Complexity             &Model Size       \\  
\midrule   
  \textbf{LAGD algorithm }       & \XSolidBrush       & $\mathcal{O}(KMN)$            & $\sim 10^{1}$                 \\
  BNN method \cite{xia2019deep}                & \checkmark         & $\mathcal{O}(NM^2+M^3)$       & $\sim 10^{3}$                 \\
  Deep Learning Method \cite{kim2020deep}      & \checkmark         & $\mathcal{O}(NM^2+M^3)$       & $\sim 10^{3}$                 \\  
  Deep unfolding WMMSE \cite{pellaco2021deep}       & \checkmark         & $\mathcal{O}(LKM^2)$          & $\sim 10^{1}$           \\     
  Adam \cite{kingma2014adam}/ GD                   & \XSolidBrush       & $\mathcal{O}(KMN)$            & \XSolidBrush             \\
  WMMSE \cite{shi2011iteratively}                      & \XSolidBrush       & $\mathcal{O}(KM^3)$           & \XSolidBrush             \\
  \bottomrule  
\end{tabular}
\end{center}
\end{table}

\textcolor{black}{Finally, we compute the complexity of various approaches in Table \ref{computational complexity}, where $K$ denotes the number of iterations of the algorithm, while $L$ is the number of inner iterations in each iteration of the algorithm \cite{pellaco2021deep}.
It can be seen that the LAGD has comparable computational complexity with the GD-based algorithm, while dramatically saving overall computational complexity due to the lack of a training stage. 
When compared with the WMMSE algorithm, the computational complexity is also significantly lower, as matrix inversion, bisection search for the Lagrange multiplier, and iterations over three variables in the WMMSE algorithm are avoided.}

\section{CONCLUSION}\label{CONCLUSION}

In this paper, we proposed a novel learning-based optimization algorithm for solving the WSR maximization problem in a downlink MISO communication system.
The proposed LAGD algorithm optimizes the transmit precoder directly based on a neural network-based GD approach. It retains the algorithmic principles of GD optimization while the performance is significantly improved thanks to the more flexible and adaptive update rule. Another important benefit of the proposed LAGD algorithm is that it is unsupervised; and hence, does not require solutions with an alternative method as training data, and is realized with a simple shallow network. The training-free implementation allows it to be used in a plug-and-play manner in different scenarios without any additional cost on model training. Through simulations, we have discovered that the higher the complexity and the non-convexity of the underlying scenario, i.e., higher SNR or more users, the superior the performance improvement obtained by the LAGD algorithm compared to WMMSE. 
In the future, we will prove the convergence of the algorithm and explore the application of the LAGD algorithm in more challenging scenarios involving channel uncertainties and multiple receive antennas.
\ifCLASSOPTIONcaptionsoff
  \newpage
\fi

\bibliographystyle{IEEEtran}
\addcontentsline{toc}{section}{\refname}\bibliography{Bib_MLBF} 

\end{document}